\documentclass[11pt]{article}
\usepackage{moriond,epsfig}

\def\ltap{\raisebox{-.6ex}{\rlap{$\,\sim\,$}} \raisebox{.4ex}{$\,<\,$}} 

\begin{document}
\vspace*{4cm}
\title{SOFT GLUON EFFECTS IN THE H$\to WW$ SEARCH CHANNEL AT THE LHC}

\author{M. GRAZZINI }

\address{INFN, Sezione di Firenze and Dipartimento di Fisica, Universit\`a di Firenze,\\ Via Sansone 1, I-50019, Sesto Fiorentino, Florence, Italy}

\maketitle\abstracts{
Non resonant production of $WW$ pairs
is the most important background in the Higgs search in
the range $155\ltap M_H \ltap 170$ GeV at the LHC.
We consider QCD corrections to
$WW$ pair production in hadron collisions and present a calculation which consistently combines NLO corrections with soft-gluon resummation at small transverse momenta of the $WW$ pair. Spin correlations are fully taken into account in our calculation.
We study the impact of the resummation on the leptonic observables and perform
a comparison with results obtained at NLO and with the MC@NLO event generator.
}

At the end of 2007 the most powerful particle collider ever built, the LHC,
is finally expected to start its operations.
One of the main issues of the LHC physics program is the search
for the Higgs boson, the agent of electroweak symmetry breaking that has so far escaped experimental discovery.
To find a Higgs boson, a signal has to be identified over a background, and, depending on the search channel,
precise predictions for the corresponding cross sections are required.
This is expecially true if, as for the decay mode $H\to WW$,
a mass peak cannot be directly reconstructed.
The $H\to WW\to l\nu l\nu$ channel is the most important in the mass range between 155 and 170 GeV.
As a consequence, the background for non resonant $WW$ production
has to be under good control.

The $WW$ cross section is presently known at next-to-leading order (NLO)
in QCD perturbation theory
\cite{Dixon:1999di,Campbell:1999ah}. The NLO effect increases the cross
section by about $40\%$ at LHC energies.

The fixed-order NLO calculations are reliable to predict $WW$
cross sections and distributions as long as the scales involved in the
process are all of the same order.  When the transverse momentum
of the $WW$ pair $p_{T}^{\rm W W}$ is much smaller than its invariant mass
$M_{\rm WW}$ the validity of the fixed-order expansion may be spoiled since
the coefficients of the perturbative expansion can be enhanced by powers of
the large logarithmic terms, $\ln^n M_{\rm WW}/p_{T}^{\rm W W}$.  This is
certainly the case for the $p_{T}^{\rm W W}$ spectrum, which, when evaluated
at fixed order, is even divergent as $p_{T}^{\rm W W}\to 0$, and thus requires
an all-order resummation of the logarithmically enhanced terms.  Resummation
effects, however, can be visible also in other observables, making it
important to study them in detail.

In the following we report on a study of soft-gluon effects in $WW$
production at hadron colliders \cite{Grazzini:2005vw}.  We use the helicity
amplitudes of \cite{Dixon:1998py} and work in the narrow width
approximation (i.e. we only consider double-resonant contributions), but fully
include the decays of the $W$ bosons, keeping track of their
polarization in the leptonic decay.  In the large $p_{T}^{\rm W W}$ region we
use LO perturbation theory ($WW$+1 parton); in the region
$p_{T}^{\rm W W}\ll M_{\rm WW}$ the large logarithmic contributions are
resummed to NLL and (almost) NNLL \cite{deFlorian:2000pr,deFlorian:2001zd}
accuracy.

To perform the resummation
we use the formalism of \cite{Bozzi:2003jy,Bozzi:2005wk}.
In this approach, the resummation is achieved at the level of the partonic
cross section and the large logarithmic contributions are exponentiated
in a process-independent manner, being constrained to give vanishing
contribution to the total cross section.
Our results have thus uniform NLO accuracy over the entire range of transverse momenta but consistently 
include the all-order resummation of logarithmically enhanced terms in the region $p_{T}^{\rm W W}\ll M_{\rm WW}$.

The results are compared with those obtained at NLO with MCFM \cite{Campbell:1999ah} and with the ones from
the general purpose event generator MC@NLO \cite{MCatNLO}, which
partially includes the effect of spin correlations
in the $W$'s decay.

To compute the $WW$ cross section we use MRST2002 NLO densities \cite{Martin:2002aw}
and $\alpha_{\mathrm{S}}$ evaluated at two-loop order.
We first consider the inclusive cross sections.
Our NLL+LO result is 115.6 pb,
and agrees with the NLO one to better than $1\%$.
The cross section from MC@NLO is instead lower, about 114.7 pb.
The above difference is due to the different choice
of the scales, and to the different convention in the choice of the electroweak couplings adopted in MC@NLO.
\begin{figure}
\begin{center}
\begin{tabular}{cc}
\epsfysize=5.5truecm
\epsffile{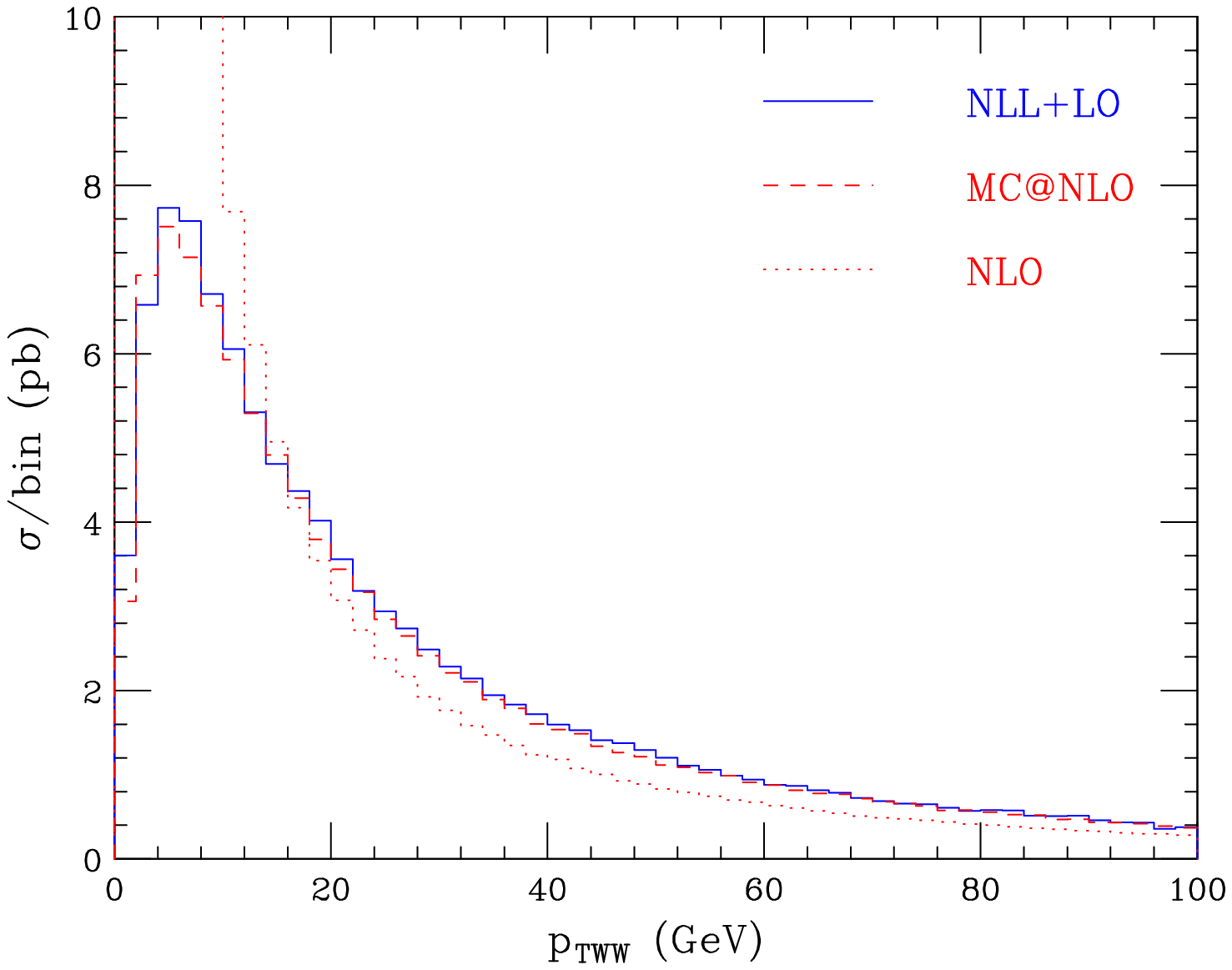} & \epsfysize=5.5truecm\epsffile{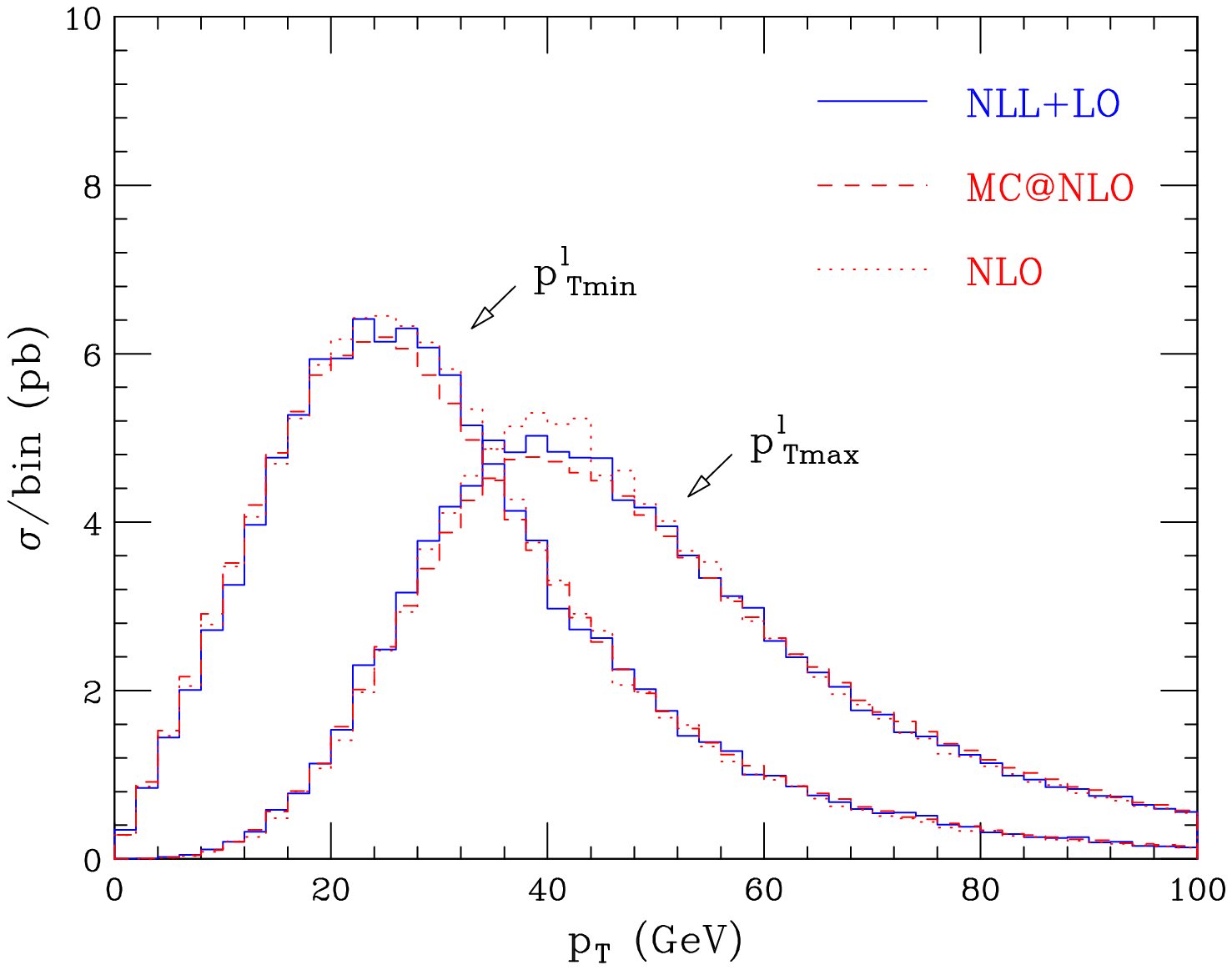}\\
\end{tabular}
\end{center}
\vspace*{-3mm}
\caption{\label{fig1}{\em Left: Comparison of the transverse momentum spectra of the ${\rm W^+W^-}$ pair obtained at NLL+LO, NLO and with MC@NLO. No cuts are applied. Right: corresponding predictions for the lepton $p_T$ spectra.}}
\end{figure}
In Fig.~\ref{fig1} (left) we show the $p_{T}^{\rm WW}$ distribution,
computed at NLO (dotted), NLL+LO (solid) and with MC@NLO (dashed).
We see that the NLO result diverges to $+\infty$ as $p_{T}^{\rm W W}\to 0$.
The NLL+LO and MC@NLO results are instead finite as $p_{T}^{\rm W W}\to 0$ and are in good agreement,
showing a kinematical peak around $p_{T}^{\rm W W}\sim 5$ GeV.

We now consider the $p_T$ spectra of the leptons.
For each event, we classify the transverse momenta of the two charged leptons
into their minimum and maximum values, $p^l_{T{\rm min}}$ and $p^l_{T{\rm max}}$.
In Fig.~\ref{fig1} (right) we plot the corresponding $p_T$ spectra,
computed at NLL+LO (solid), NLO (dotted) and with MC@NLO (dashes).
All the three predictions are clearly in good agreement:
the effect of resummation, which is essential in the $p_{T}^{\rm W W}$ spectrum,
is hardly visible in the leptonic spectra.

To further assess the effect of resummation,
we consider the application of the following cuts,
suggested by the study of \cite{Davatz:2004zg}:

\begin{itemize}
\item For each event, $p^l_{T{\rm min}}$ should be larger than $25$ GeV and $p^l_{T{\rm max}}$ should be between $35$ and $50$ GeV.
\item The invariant mass $m_{ll}$ of the charged leptons should be smaller than $35$ GeV.
\item The missing $p_T$ of the event should be larger than $20$ GeV.
\item The azimuthal charged lepton separation in the transverse plane $\Delta\phi$ should be smaller than $45^o$.
\item A jet veto is mimicked
by imposing that the transverse momentum of the $WW$ pair should be smaller than $30$ GeV.
This cut is perfectly legitimate in our resummed calculation and is exactly equivalent to a jet veto at NLO.
\end{itemize}
These cuts, designed for the search of a Higgs boson with $M_H = 165$ GeV,
strongly select the small $\Delta\phi$ region. The jet veto is used
in order to reduce
the $t{\bar t}$ contribution, which is expected to produce large-$p_T$ $b$-jets from the decay of the top quark.

The NLL+LO (MC@NLO) accepted cross section is 0.599 pb (0.570 pb) which should be contrasted with the NLO result,
which is 0.691 pb, about $20\%$ higher.
This relative large difference is due to the fact that these cuts
enhance the relevance of the small-$p_{T}^{\rm W W}$ region,
where the NLO calculation is not reliable.

\begin{figure}[htb]
\begin{center}
\begin{tabular}{c}
\epsfysize=7truecm
\epsffile{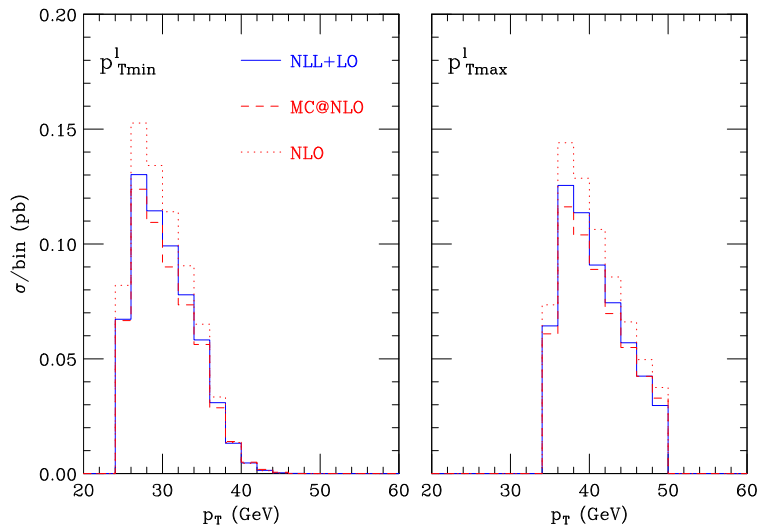}\\
\end{tabular}
\end{center}
\vspace*{-3mm}
\caption{\label{WWfig2}{\em Distributions of $p^l_{T{\rm min}}$ and $p^l_{T{\rm max}}$ when cuts are applied.}}
\end{figure}

In Fig.~\ref{WWfig2} the $p^l_{T{\rm min}}$ and $p^l_{T{\rm max}}$ distributions are presented.
We see that although the three predictions are
in reasonable agreement in shape,
differences are now evident.
In particular, the $p^l_{T{\rm min}}$ distribution at NLO is steeper than the other two.
Comparing NLL+LO and MC@NLO spectra, we see that
the former are steeper than the latter:
with the application of strong cuts
the differences between NLL+LO and MC@NLO predictions are enhanced.

\begin{figure}[htb]
\begin{center}
\begin{tabular}{cc}
\epsfysize=5.5truecm
\epsffile{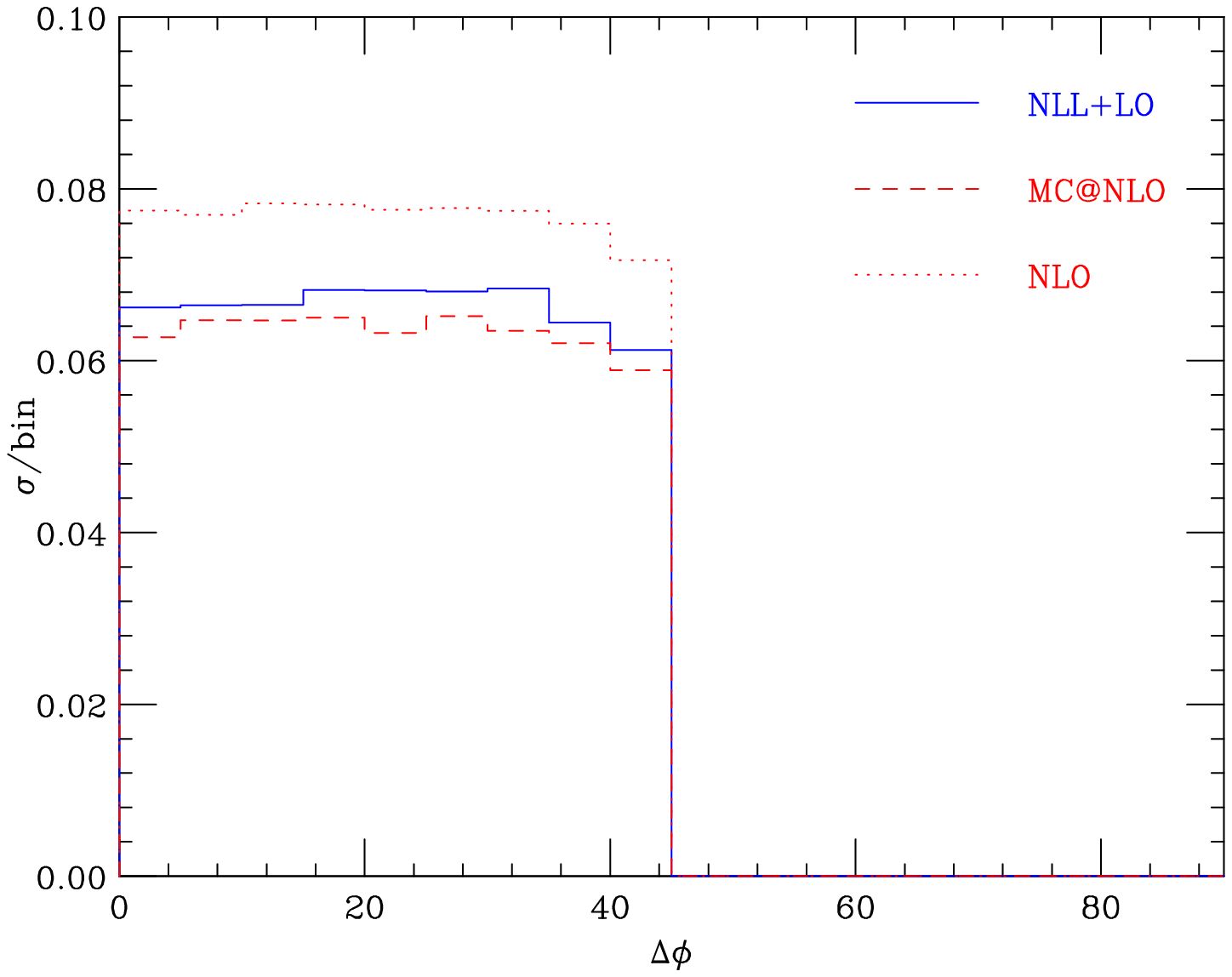} &\epsfysize=5.5truecm\epsffile{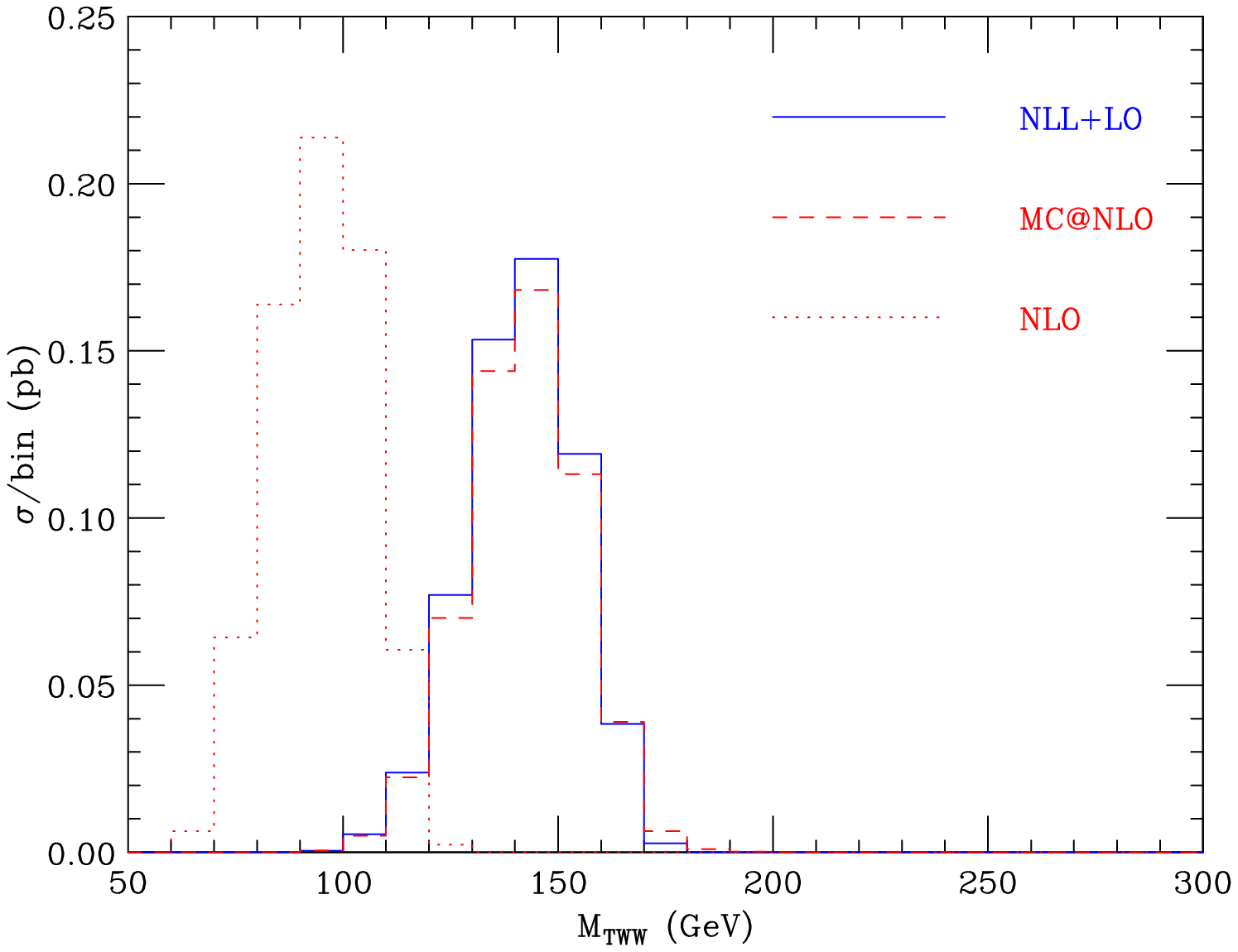}\\
\end{tabular}
\end{center}
\vspace*{-3mm}
\caption{\label{fig3}{\em Left: $\Delta\phi$ distribution when cuts are applied. Right: Transverse-mass distribution. }}
\end{figure}
In the search for the Higgs boson in the $H \to WW \to l\nu l\nu$ channel
an important difference between the signal and the background is found in the $\Delta\phi$ distribution.
Since the Higgs is a scalar
the signal is expected to be peaked at small values of $\Delta\phi$ \cite{Dittmar:1996ss},
whereas the $\Delta\phi$ distribution for the background is expected to be reasonably flat.
It is thus important to assess the effect of resummation on this distribution,
which is also known to be particularly sensitive to spin correlations.
In Fig.~\ref{fig3} (left) the $\Delta\phi$ distribution is displayed.
We see that the shapes of the three results
are in good agreement with each other, although
a slightly different slope of the
NLL+LO result with respect to MC@NLO and NLO appears.
We remind the reader that the NLO and NLL+LO calculations exactly include spin correlations, whereas
MC@NLO neglects spin correlations in the finite (non-factorized) part of the one-loop contribution.

In Fig.~\ref{fig3} (right) we finally consider the transverse-mass distribution of the $WW$ system, defined as in \cite{Rainwater:1999sd}.
The NLO result (dotted) is compared to the NLL+LO one (solid) and to MC@NLO (dashes).
We see that the effects of soft-gluon resummation are dramatic for this distribution:
the NLL+LO result is shifted towards larger values
of $M_{TWW}$ by about 50 GeV.
We find that this big difference is
mainly due to the leptonic cuts: removing the jet veto the shift in
the transverse-mass distribution is basically unchanged.
Comparing the shapes of the histograms
we see that at NLO the shape
is fairly different with respect to NLL+LO and MC@NLO.
Also the NLL+LO and MC@NLO distributions now show clear differences:
the position of the peak is the same, but
the NLL+LO result is steeper and softer than the MC@NLO one.

In this talk we have discussed soft-gluon effects
in $WW$ production at the LHC.
We showed that resummation has a mild impact on inclusive leptonic distributions.
On the other hand, when stringent cuts are applied,
the effects of resummation are strongly enhanced.
The most significant effect is seen in the transverse mass distribution, for which the NLO calculation
is clearly not reliable.
Our resummed predictions are generally in good agreement with those of the MC@NLO event generator.

\end{document}